\documentclass[preprint,endfloats,pre,draft]{revtex4}
\usepackage{bm}
\usepackage[final]{graphicx}
\usepackage{longtable}
\newcommand{\be}{\begin{equation}}
\newcommand{\ee}{\end{equation}}
\newcommand{\bea}{\begin{eqnarray}}
\newcommand{\eea}{\end{eqnarray}}

\begin{document}

\title{Field theoretical analysis of adsorption of polymer chains at surfaces:Critical exponents and Scaling.}

\author{Z. Usatenko$^{1}$}

\affiliation{$^{1}$Institute for Condensed Matter Physics, National
Academy of Sciences of Ukraine, 1~Svientsitskii Str., UA--79011
Lviv, Ukraine } \email{pylyp@ph.icmp.lviv.ua}

\date{\today}

%\vspace{0.1cm}

\begin{abstract}

The process of adsorption on a planar repulsive, "marginal" and
attractive wall of long-flexible polymer chains with excluded volume
interactions is investigated. The performed scaling analysis is
based on formal analogy between the polymer adsorption problem and
the equivalent problem of critical phenomena in the semi-infinite
$|\phi|^4$ n-vector model (in the limit $n\to 0$) with a planar
boundary. The whole set of surface critical exponents characterizing
the process of adsorption of long-flexible polymer chains at the
surface is obtained. The polymer linear dimensions parallel and
perpendicular to the surface and the corresponding partition
functions as well as the behavior of monomer density profiles and
the fraction of adsorbed monomers at the surface and in the interior
are studied on the basis of renormalization group field theoretical
approach directly in $d=3$ dimensions up to two-loop order for the
semi-infinite $|\phi|^4$ n-vector model. The obtained field-
theoretical results at fixed dimensions $d=3$ are in good agreement
with recent Monte Carlo calculations. Besides, we have performed the
scaling analysis of center-adsorbed star polymer chains with $f$
arms of the same length and we have obtained the set of critical
exponents for such system at fixed $d=3$ dimensions up to two-loop
order.

\end{abstract}
\maketitle
\section{Introduction}
The biologically relevant systems such as polymers, membranes,
 thin films etc.  exhibit a rich variety of phase transitions,
 and the theory of critical phenomena in such complex systems is far
 from complete. From a practical point of view
adsorption phenomena in polymeric solutions are important for such
processes as lubrication, adhesion and surface protection
\cite{Grassberger,Zhao}, as well as biological processes of membrane
- polymer interaction. The statistical properties of the adsorption
of long, flexible macromolecular chains (polymers) at surfaces in
dilute, semi-dilute, and concentrated polymer solutions have found
considerable interest.

Long flexible polymer chains immersed into a good solvent are very
well described by the model of self-avoiding walks (SAWs) on a
regular lattice \cite{Cloizeaux}. The SAW exhibits critical behavior
approaching the limit of infinite number of steps which can be
extracted from the $n\to 0$ limit of an $O(n)$ symmetric field
theory\cite{deGennes}.

The average square end-to-end distance, the number of configurations
with one end fixed and with both ends fixed at the distance
$x=\sqrt{({\vec x}_{A}-{\vec x}_{B})^{2}}$ exhibit the following
asymptotic behavior in the limit $N\to \infty$ \be <R^2>\sim
N^{2\nu},\quad\quad\quad Z_{N}\sim q^{N}N^{\gamma-1},\quad\quad\quad
Z_{N}(x)\sim q^{N}N^{-(2-\alpha)},\label{RZ} \ee
 respectively.  Exponents $\nu$, $\gamma$ and $\alpha$ are the universal
 correlation length, susceptibility and specific heat critical
 exponents for the $n=0$ model, $d$ is the space dimensionality, $q$
 is a non universal fugacity. The value $1/N$ plays a role of a critical
 parameter  analogous to the reduced  critical temperature in
 magnetic systems.

 The critical behavior of a system is strongly affected by the
presence of a confining surface.  As noted by de Gennes
\cite{deGennes,deGennes1} and by Barber et al. \cite{Barber}, there
is a formal analogy of the polymer adsorption problem to the
equivalent problem of critical phenomena in the semi-infinite
$|\phi|^4$ $n$-vector model of a magnet with a free surface
\cite{DD81,DD83,D86}. In this case any bulk universality class is
split into several surface universality classes, with new surface
critical exponents \cite{DD81,DD83,D86}. It should be mentioned that
such analogy takes place for a real polymer chain, i.e., polymers
with excluded volume interaction.

Based on the above analogy, the problem of adsorption of
long-flexible  linear polymer chains on a surface in the case of
pure solvent was investigated few years ago by Eisenriegler and
co-workers \cite{EKB82,eisenriegler:83,Eisenriegler}. This theme was
a subject of a series of works (for the sake of brevity we mention
only few of them \cite{deGennes,HG94,SKG99,SGK01,RDGKS02,ZLB90}).
The series of works \cite{Ochno1,Ochno2} was dedicated to
theoretical and numerical investigation of statistical properties of
the star polymers in a good solvent in the dilute limit. Our recent
work \cite{Usatenko} was dedicated to the field-theoretic
investigation of the influence of different kinds of bulk disorder
in a polymeric solution on the process of adsorption of
long-flexible polymers at the surface.

The large discrepancies between Monte Carlo results and
$\epsilon$-expansion estimates in  Ref.\cite{EKB82} of some critical
exponents characterizing the process of adsorption at the surface of
long-flexible polymer chains motivated us for the present work.

In this paper, we investigate the adsorption phenomena of
long-flexible linear polymer chains at repulsive, "marginal" and
attractive surface in sufficiently dilute pure polymeric solution,
so that interchains interactions are neglected. Thus, for collecting
the full information about the process of adsorption of
long-flexible polymer chains on a surface is sufficient to consider
surface effects for configurations of a single chain. The proposed
technic gives also possibility to perform the scaling analysis of
the statistical properties of the center-adsorbed star polymers with
 $f$ arms which have the same length in dilute solutions.

 At sufficiently low temperatures, $T<T_a$, the
 attraction between the monomers and the surface leads to an adsorbed
 state, where a finite fraction of the monomers is attached to the
 system boundary.
 The deviation from the adsorption threshold,
 $c\propto(T-T_a)/T_a)$, changes sign at the transition between the adsorbed
 ($c<0$) and the nonadsorbed state ($c>0$) and it plays a role of a
 second critical parameter. The adsorption threshold for infinite
 polymer chains, where $1/N\to 0$ and $c=c_{0}^{sp}\to 0$ is a multicritical
 phenomenon. Thus, the ordinary transition $(T>T_a)$ corresponds to
 a repulsive surface, the multicritical point  corresponds
 to a "marginal" surface at $T=T_a$ and for the attractive surface $T<T_a$
the $d-1$ dimensional criticality at the "surface transition" line
takes place. As it known \cite{DD81,DD83,DSh98}, for each of these
transitions the knowledge of one independent surface critical
exponent, for example critical exponent $\eta_{\parallel}$ of
correlations in directions parallel to the surface, give access to
the whole set of the other surface critical exponents via surface
scaling relations and the bulk critical exponents $\nu$ and $\eta$.
The crossover critical exponent $\Phi$ characterizes the crossover
behavior between the special and ordinary transitions ($c\ne 0$).
The latter exponent is related to the length scale
\cite{eisenriegler:83,Eisenriegler}, \be \label{xic} \xi_c \sim
|c|^{-\nu/\Phi}, \ee
 associated with the parameter $c$.  In the polymer problem the length
 $\xi_c$ can be interpreted as the distance from the surface up to
 which the properties of the polymer chains depend on the value of $c$, not
 only on its sign.  The remaining, bulk length scales are the average
 end-to-end distance $\xi_R=\sqrt{<R^2>}\sim N^{\nu}$ and the
 microscopic length $l$ -- the effective monomer linear
 dimension. Near the multicritical point the only relevant lengths are
 $\xi_R\to\infty$ and $\xi_c\to\infty$. Correspondingly, the properties of the
 system depend on the ratio $\xi_R/\xi_c$. In the asymptotic scaling
 regime the universal physical quantities $X(N,c)$ and $Y(z;N,c)$
 assume the scaling forms
\be \label{scal}
X(N,c)=N^{a_X}X^s_{\pm}\big(\xi_R/\xi_c\big),\hskip1cm
Y(z;N,c)=N^{a_Y}Y^s_{\pm}\big(z/\xi_R,\xi_R/\xi_c\big), \ee %%
where $X^s_{\pm}$ and $Y^s_{\pm}$ denote the scaling functions with
the subscripts $+$ and $-$ corresponding to $c>0$ and $c<0$
respectively.  The characteristic length ratio is
$(\xi_R/\xi_c)^{\Phi/\nu} \sim |c|N^{\Phi}$, where $cN^{\Phi}$ is
the standard scaling variable \cite{EKB82}.  The exponents $a_X$ and
$a_Y$  assume different values for different quantities $X$ and $Y$.
The scaling analysis of different important characteristics of
process of adsorption of long-flexible linear polymer chains and
 center-adsorbed star polymers with the same arm length $f$ in a good solvent in the dilute limit will be
discussed in the next sections.

\section{the model}
The presence of a hard wall leads to a modification of the
interactions in the near-surface layer. Thus, in the semi-infinite
system there should be additional, surface contribution to the
Hamiltonian. The effective Hamiltonian of the semi-infinite
$|\phi|^4$ $n$-vector model is \cite{DSh98}

\bea H_{eff} & = & \int_{V} d^{d}x [\frac{1}{2} \mid
\nabla\vec{\phi} \mid ^{2} + \frac{1}{2} \mu_{0}^{2}\vec{\phi}^{2}
+\frac{1}{4!} v_{0}
(\vec{\phi}^2)^{2}\nonumber\\
& + &\frac{c_{0}}{2} \int_{\partial V}d^{d-1}r
\vec{\phi}^{2}({\bf{r}},z=0),\label{9} \eea
where $\vec{\phi}(x)$ is an $n$-vector field with the components
$\phi_{i}(x)$, $i=1,...,n$. Here $\mu_{0}^2$ is the "bare mass",
which in the case of a magnet corresponds to the reduced
temperature. The limit $n\to 0$ of this model at its critical point
and in the limit of an infinite number of steps $N$ can be
interpreted as a model of long-flexible polymer chains near the
surface. It should be mentioned that the $d$-dimensional spatial
integration is extended over a half-space $I\!\!R^d_+\equiv\{{\bf
x}{=}({\bf r},z)\in I\!\!R^d\mid {\bf r}\in I\!\!R^{d-1}, z\ge 0\}$
bounded by a plane free surface at $z=0$. The fields $\phi_{i}({\bf
r},z)$ satisfy the Dirichlet boundary condition in the case of
ordinary transition: $\phi_{i}({\bf r},z)=0$ at $z=0$ and the
Neumann boundary condition in the case of special transition:
$\partial_{z}\phi_{i}({\bf r}, z)=0$ at $z=0$ \cite{DD81,DD83}.

The value $c_{0}$ describes the surface-enhancement of interactions.
In the polymer analog $c_0\propto (T-T_a)/T_a$, as already noted in
the introduction. The cases $c>0$, $c=0$ and $c<0$ correspond,
respectively, to the repulsive, marginal and attractive surfaces.
The surface introduces an anisotropy into the problem, and
directions parallel and perpendicular to the surface are no longer
equivalent. In accordance with the fact that we have to deal with
semi-infinite geometry $({\bf x}=({\bf r},z\geq 0))$, only parallel
Fourier transforms in $d-1$ dimensions can be performed. We shall
denote the associated parallel momenta as $p$.

Besides, we shortly consider the case of center-adsorbed star
polymers with $f$ arms of the same length at a free surface. The
corresponding Landau-Ginzburg-Wilson Hamiltonian of $f$ interacting
fields $\phi_{\alpha}$ with $n$ components in semi-infinite geometry
reads
\begin{eqnarray}
 H_{eff} & = & \sum_{\alpha=1}^{f}\int_{V} d^{d}x [\frac{1}{2}
\mid \nabla\vec{\phi}_{\alpha} \mid ^{2} + \frac{1}{2}
\mu_{0}^{2}\vec{\phi}_{\alpha}^{2}] +
\sum_{\alpha,\beta=1}^{f}u_{\alpha,\beta}\int d^{d}x
\vec{\phi}_{\alpha}^{2}(x)
\vec{\phi}_{\beta}^{2}(x)\nonumber\\
& + &\frac{c_{0}}{2} \int_{\partial V}d^{d-1}r
\vec{\phi}^{2}({\bf{r}},z=0),\label{10} \end{eqnarray} where the
value $f$ corresponds to number of arms of center-adsorbed star
polymers.

\section{Critical behavior of long-flexible polymer chains near the marginal surface
for $c=c_{0}^{sp}$ (special transition) }

In our investigation we use the scheme of investigation of
semi-infinite $|\phi|^4$ $n$-vector model, proposed in work
\cite{DSh98} and partially described for the case of long-flexible
polymer chains in \cite{Usatenko}. As was indicated in
\cite{Usatenko}, and easy to see from results in \cite{DSh98}, in
order to obtain the universal surface critical exponents,
characterizing the adsorption at a wall of long-flexible polymer
chains in good solutions, it is sufficient to consider the
correlation function of two surface fields $G^{(0,2)}$.  The
universal surface critical exponents for such systems depend on the
dimensionality of space $d$ and the number of order parameter
components $n (n\to 0)$.

In order to investigate the crossover scaling behavior from the
nonadsorbed region, $c>c_{0}^{sp}$, to the adsorbed one,
$c<c_{0}^{sp}$, let us consider a small deviation $\Delta
c_{0}=c_{0}-c_{0}^{sp}$ from the multicritical point. Using the
above mentioned scheme, we obtain the following asymptotic scaling
form of the surface correlation functions of the long-flexible
polymer chains with one or two ends fixed at the surface, \be
G^{\lambda}(z, c_{0})\sim z^{1-\eta^{sp}_{\perp}}G_{\lambda}(\tau
z^{1/\nu}, \tau^{-\Phi} \Delta c_{0})\ee
\begin{equation}
G_{\parallel,\perp}(r;\mu_{0},u_{0},c_{0})\sim
r^{-(d-2+\eta_{\parallel,\perp}^{sp})}G_{\parallel,\perp}(\tau
r^{1/\nu};\tau^{-\Phi}\Delta c_{0}), \label{29b}
\end{equation}

where $\eta_{\parallel}^{sp}=\eta_{1}^{sp}+\eta $ is the surface
exponents at the multicritical point which characterizes the
critical point correlations parallel to the surface and
$\eta_{\perp}=\frac{\eta+\eta_{\parallel}}{2}$ is the surface
critical exponents which characterizes the critical point
correlations perpendicular to the surface; $\nu$, $\eta$ are the
usual bulk exponents.

We also obtain for $\Delta c$ and for the scaling variable $\bar{c}$
the following asymptotic forms
\begin{equation}
\Delta c\sim \mu^{-\eta_{\bar{c}}(u^{*})} \Delta c_{0},\quad\quad
\Delta c\sim \tau^{-\nu \eta_{\bar{c}}(u^{*})} \Delta
c_{0}\label{28}
\end{equation} and
\begin{equation}
\bar{c}\sim \mu^{-(1+\eta_{\bar{c}}(u^{*}))} \Delta
c_{0},\quad\quad\quad \bar{c}\sim \tau^{-\Phi} \Delta
c_{0},\label{29}
\end{equation}
where
\begin{equation}
\Phi=\nu (1+\eta_{\bar{c}}(u^{*})) \label{30}
\end{equation}
is the surface crossover critical exponent
\cite{DSh98,UH03,Usatenko}. Eq. (\ref{29}) explains the physical
meaning of the surface crossover exponent as a value which
characterizes the measure of deviation from the multicritical point.

The series expansions for the critical exponent $\eta_{\parallel}$
and $\eta_{\bar{c}}$ up to two-loop order was obtained previously
for the pure case in \cite{DSh98} and for the case with some amount
of disorder in \cite{UH02,UH03}(the pure case can be obtained from
the formulas (4.3) in \cite{UH02} and (6.13) in \cite{UH03} in the
limit $n\to 0$ and at $v=0$). Thus, we have
\begin{equation}
\!\!\eta_\|^{sp} =-\frac{u}{8}+\frac{3}{8}A(0)u^2,\label{37}
\end{equation}
where
\begin{equation}
A(0)=2A- \frac{5}{24}+\frac{ln{2}}{3}(ln 2 - 1),\label{36}
\end{equation}
and
 \be \eta_{\bar{c}}=-\frac{1}{2}(ln2-\frac{1}{4})u-
\frac{3}{4}D(0)u^2,\label{48} \ee
where
 \be
D(0)=A-B+ln{2}(\frac{2}{3}ln{2}-\frac{1}{2})+\frac{11}{48},\label{47}
\ee where $A=0.202428$, $B=0.678061$ are integrals originating from
the two-loop melon-like diagrams (see \cite{DSh98}). The knowledge
of above mentioned surface critical exponents $\eta_{\parallel}$ and
$\Phi$ and bulk critical exponents $\nu$ and $\eta$ give us
possibility to perform the scaling analysis of different
characteristics near the multicritical point $1/N\to 0$ and $c\to
0$. So, it allow us to investigate the rich and interesting critical
behavior of long-flexible polymer chains near the  marginal and
adsorbing surface. Let us first consider the mean square end-to-end
distance for one end attached to the surface and the other one free.
In the semi-infinite system the translational invariance is broken,
and the parallel $ <R^2_{\parallel}>$ and perpendicular
$<R^2_{\perp}>$ parts of the average end-to-end distance
$<R^2>=<R^2_{\perp}+R^2_{\parallel}>$
 should be distinguished. For $<R^2_{\perp}>^{1/2}$ the exponent in
 the scaling form (\ref{scal}) is $a_X=\nu$ and  the corresponding scaling
 functions assume the form $\sim const$ for $c\ge 0$ and
 $\sim 1/y$
 for $c< 0$, where $y=\xi_R/\xi_c$ \cite{EKB82}.  Thus, for the
 adsorbed state and for $N\to\infty$ the length associated with $c$
 describes the thickness $\xi$ of the adsorbed layer,

\be \label{xi} \xi = <R^2_{\perp}>^{1/2}\sim\xi_c \hskip1cm c<0. \ee
This thickness diverges for $c=1/N=0$ and for finite negative values
of $c$ remains finite for an infinite chain. For $c\ge 0$ the
asymptotic behavior of the mean distance of the free end from the
other end attached to the surface is \be \label{xi>0}
 <R^2_{\perp}>^{1/2}\sim N^{\nu}\hskip1cm c \ge 0,
\ee
 i.e it has the same asymptotic behavior as in the bulk. The
 asymptotic scaling form of $ <R^2_{\parallel}>^{1/2}$ for $c<0$ is
 $<R^2_{\parallel}>^{1/2}\sim
 |c|^{(\nu^{d-1}-\nu)/\Phi}N^{\nu^{d-1}}$, where $\nu^{d-1}$ is the
 correlation exponent in $d-1$ dimensions. For $c\ge 0$ the scaling
 form of $<R^2_{\parallel}>^{1/2}$ is given by Eq.  (\ref{xi>0})),
 i.e. it is also the same as in the bulk. The one-loop calculations for
 this exponent based on renormalization group (RG) field theory at fixed dimensions $d=3$ gives
  $(\nu^{d-1}-\nu)/\Phi=0.335$ and two-loop result is $0.313$. The
  comparison of the obtained results with the last results of Monte-Carlo simulations \cite{HG94}
  is presented in Fig.1 for the case $c<0$.

  \begin{figure}[htb]
\begin{center}
\includegraphics[scale=0.8]{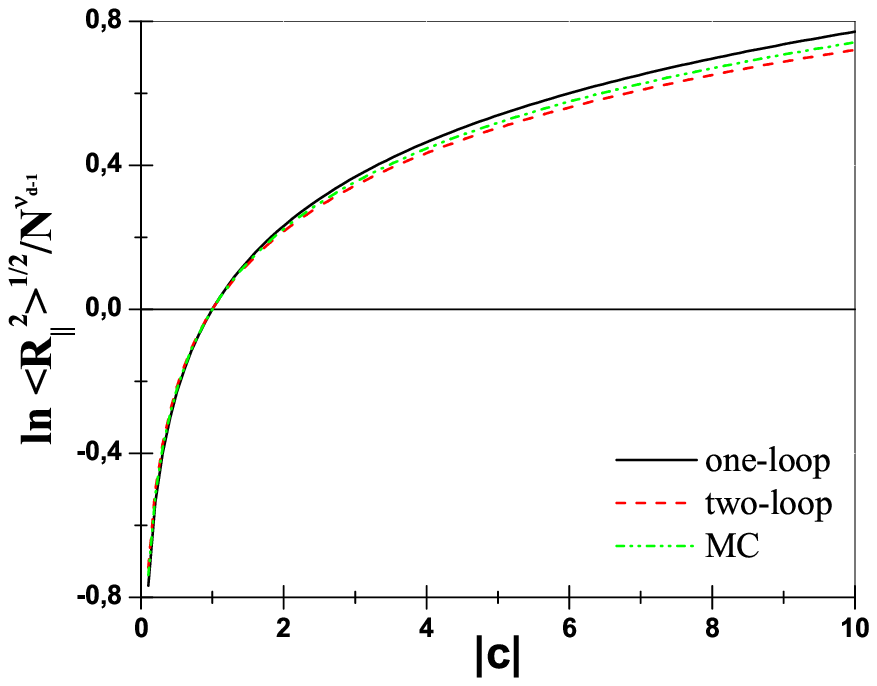}
\end{center}
\caption{Color online. The asymptotic scaling form of the parallel
part of the end-to-end distance $ ln
<R^2_{\parallel}>^{1/2}/N^{\nu_{d-1}}$ for $c<0$ as function of
$|c|$,where $(c\propto (T-T_a)/T_a)$.}\label{fig1}
\end{figure}

As was mentioned in \cite{EKB82}, the knowledge of (\ref{9}) and
(\ref{10}) gives access to the short-distance behavior for $1\ll
 z/l,r/l\ll N^{\nu}$ at the threshold of the corresponding partition functions with one
end fixed and another end free and for the partition function with
two ends at the surface
\begin{equation}
Z^{\lambda}(0,z)\sim z^{a_{\lambda}} N^{b_{\lambda}},\label{Zl}
\end{equation}
\begin{equation}
Z^{\parallel,\perp}(r)\sim r^{a_{\parallel,\perp}}
N^{b_{\parallel,\perp}},\label{Zpp}
\end{equation}
where critical exponents are: $a_{\lambda}=
\eta_{\parallel}-\eta_{\perp}$, $b_{\lambda}=-1+\gamma_{\parallel}$
and  $a_{\parallel,\perp}=1-\eta_{\parallel,\perp}-\Phi/\nu$,
$b_{\parallel,\perp}=-1-\nu (d-1)+\Phi$. The whole set of surface
critical exponent can be obtained from the scaling relations (see
\cite{DSh98}) on the basis of series expansions for
$\eta_{\parallel}$, $\eta_{\bar{c}}$ and series for usual bulk
critical exponents $\nu$ and $\eta$.  Table 1 represents the
obtained one-loop and two-loop calculations for the corresponding
critical exponents characterizing the process of adsorption of
long-flexible polymer chains on the marginal surface (the case of
adsorption threshold).

\begin{table}[htb]
\caption{\label{tab:tab1}Critical exponents characterizing scaling
critical behavior of the long-flexible polymer at the adsorption
threshold $c=c_{0}^{a}$ (the case of marginal surface) for $d=3$ up
to one-loop and two-loop order calculated at the pure fixed point
$u^{*}=1.632$.}
\begin{center}
\begin{tabular}{rrrrrrrrrrrr}
\hline $    $~&~$ a_{\lambda} $~&~$ b_{\lambda} $~&~$ a_{\parallel}
$~&~$ b_{\parallel} $~&~$ a_{\perp} $~&~$ b_{\perp} $~&~$ a^{'}
$~&~$ b^{'}$~&~$ \bar{a} $~&~$ \bar{b} $~&~$ -\nu/\Phi $ \\
\hline one-loop & -0.079 & -0.270  & 0.490 & -1.792 &
0.411 & -1.792 & -0.094 & 0.299 & 0.053 & 0.723 & -1.434 \\
\hline
 two-loop &  -0.080  &  -0.334  &  0.252  & -1.658  &  0.172  & -1.658 &
 -0.079
 & 0.207 &  0.180  &  0.725 & -1.135  \\

\end{tabular}
\end{center}
\end{table}

We also present the  comparison of obtained in the frames of RG
field theory results at fixed dimension $d=3$ and results of the
recent Monte Carlo calculations and $\epsilon$- expansion results in
the Figures 2-4. It should be mentioned, that in the frames of RG
field theory at fixed dimensions $d=3$ we obtain the correct
positive sign for the critical exponent $a_{\perp}$.  The previous
result for this exponent on the basis of $\epsilon$-expansions was
equal $-0.09$. So, in the previous theoretical results obtained on
the basis of $\epsilon$ -expansion were large discrepancies with
Monte Carlo results $a_{\perp}=0.45$ \cite{EKB82}, $a_{\perp}=0.314$
\cite{ML88},  $a_{\perp}=0.244$ \cite{HG94}. We obtain for this
critical exponent results $a_{\perp}=0.411$ (one-loop) and
$a_{\perp}=0.172$ (two-loop), which are in good agreement with
recent results of numerical calculations \cite{ML88,HG94}. Besides,
we obtain for the critical exponent $a_{\parallel}=0.490$ (one-loop)
and $a_{\parallel}=0.252$ (two-loop). These results are in good
agreement with Monte Carlo results $0.466$ \cite{EKB82} and $0.371$.
Note, that the previous theoretical result obtained on the basis of
$\epsilon$-expansion was equal $a_{\parallel}=0.02$. \cite{EKB82}.

\begin{figure}[htb]
\begin{center}
\includegraphics[scale=0.8]{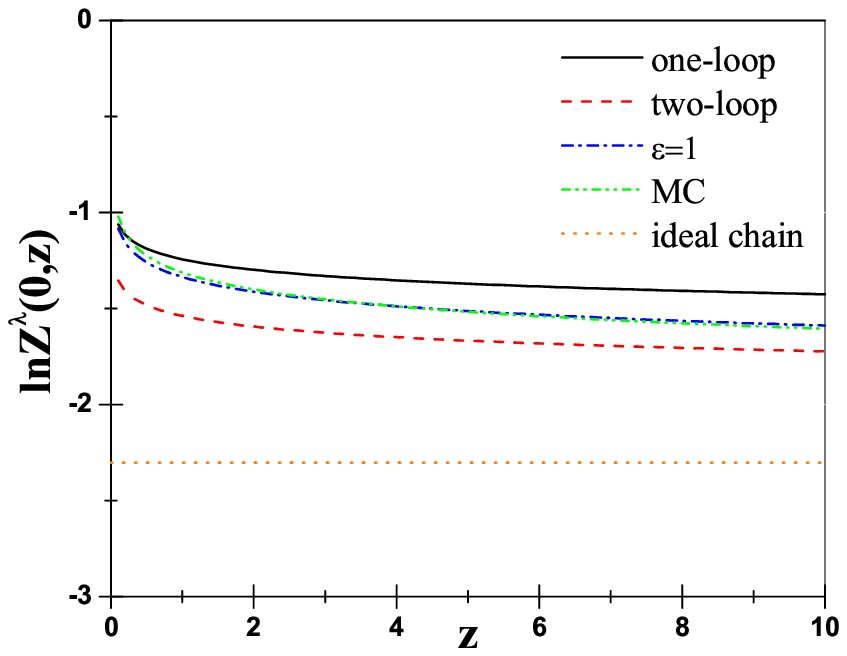}
\end{center}
\caption{Color online. The  partition function $ln Z^{\lambda}(0,z)$
just at the adsorption
 thereshold $c=0$ and for $N=100$, as a function of $z/l$ for $1\ll
 z/l\ll N^{\nu}$.  $Z^{\lambda}(z)$ is
 dimensionless and $l$ is the microscopic length scale.
 }
 \label{fig2}
\end{figure}
\begin{figure}[htb]
 \begin{center}
 \includegraphics[scale=0.8]{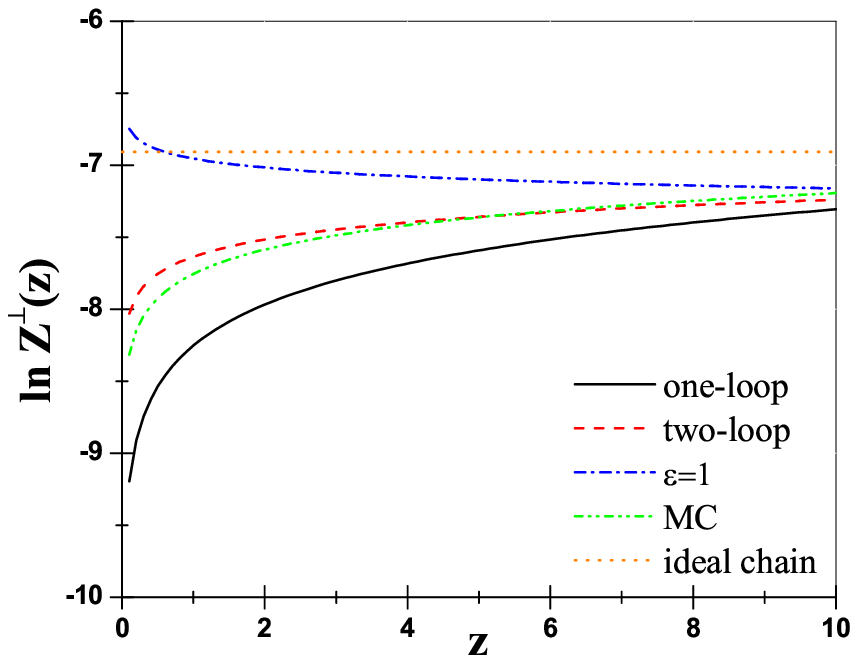}
\end{center}
\caption{Color online. The  partition function $ln Z^{\perp}(z)$
just at the adsorption
 thereshold $c=0$ and for $N=100$, as a function of $z/l$ for $1\ll
 z/l\ll N^{\nu}$.  $Z^{\perp}(z)$ is
 dimensionless and $l$ is the microscopic length scale.
 }
 \label{fig3}
\end{figure}
\begin{figure}[htb]
 \begin{center}
 \includegraphics[scale=0.8]{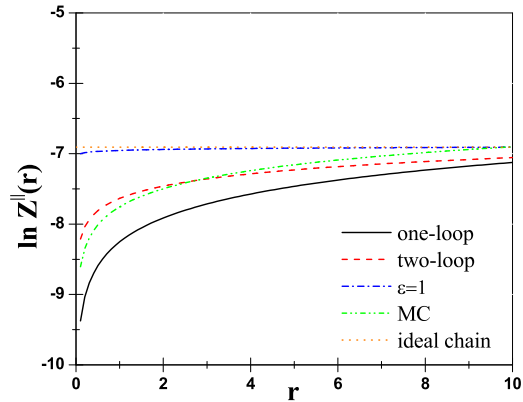}
\end{center}
\caption{Color online. The  partition function $ln Z^{\parallel}(r)$
just at the adsorption
 thereshold $c=0$ and for $N=100$, as a function of $r/l$ for $1\ll
 r/l\ll N^{\nu}$.  $Z^{\parallel}(r)$ is
 dimensionless and $l$ is the microscopic length scale.
 }
 \label{fig4}
\end{figure}

 For the fraction of monomers at the surface, $N_1/N$, the
following asymptotic behavior has been found for $N\to\infty$
\cite{EKB82,Eisenriegler}, %%
 \begin{eqnarray}
\label{N1/N} N_{1}/N\sim \left\{
 \begin{array}{lll}
|c|^{(1-\Phi)/\Phi} &\;\; \mbox{if}  & \;\;\; \mbox{$c<0$}\\
N^{\Phi-1} &\;\; \mbox{if}& \;\;\;\mbox{$c=0$}\\
(cN)^{-1} &\;\; \mbox{if}& \;\;\;\mbox{$c>0$}
 \end{array}
 \right. .
 \end{eqnarray}
 Hence, for $N\to\infty$ and for finite, negative values of $c$,
 $N_{1}/N$ is finite, but for $c\ge 0$ $N_{1}/N\to 0$ for
 $N\to\infty$.
The thickness of the adsorbed layer $\xi$ is closely related to the
 fraction of monomers at the surface $N_1/N$ \cite{EKB82,Eisenriegler}, since
 the more monomers are fixed at the wall, the smaller the region
 occupied by the remaining monomers. In the frames of one-loop approximation we obtain
$(1-\Phi)/\Phi=1.358$ and two-loop result is 0.931. In particular,
for weakly adsorbed phase $ c \leq 0$ we find
$N_{1}/N\sim\xi^{-(1-\Phi)/\nu}$. We have obtained the corresponding
series expansion for this critical exponent

\be \frac{1-\Phi}{\nu}= 1-\frac{2}{3}(\frac{3}{4}-ln
2)\bar{u}+\frac{2}{3}F(0)\bar{u}^2,\label{phn} \ee where \be
F(0)=\frac{389}{648}+2(A-B)+ln 2 (\frac{4}{3}ln 2-1), \ee

and coupling constant $\bar{u}$ is normalized in standard fashion
$\bar{u}=\frac{3}{4}u$. The calculation are performed at the
standard fixed point $u^{*}=1.632$, obtained from the Pade-Borel
resummation of beta functions $\beta(u)$ at two-loop order
\cite{Parisi}.
 Besides, we have performed Pade-analysis of the obtained
series. The result can be find in Table 2. The one-loop
($(1-\Phi)/\nu=0.947$) and two-loop (0.820) results, which we can
obtain from direct substitution of results obtained in Ref.
\cite{DSh98} are different from results obtained on the basis of
Pade-analysis. For the more reliable estimation of this critical
exponent we accept the mean value of $[2/0]$ and $[0/2]$
Pade-approximants equal to 0.620.
 The comparison of
obtained results with recent Monte Carlo calculations and results of
$\epsilon$-expansion at $d=3$ \cite{EKB82} is presented in  Figure
5.  Figure 6 contains the dependence of the thickness of the
adsorbed layer $\xi/l$ on $c<0$, where $c\propto(T-T_a)/T_a)$ is the
reduced temperature distance from the threshold. As it is easy to
see from these figures, the RG field theory directly at fixed
dimension $d=3$ \cite{DSh98,Usatenko} gives more reliable results
than field theory based on $\epsilon$-expansion
\cite{EKB82,DD81,DD83}. We see from scaling analysis, that obtained
results more better consist with resent Monte Carlo calculations
\cite{ML88,HG94}.

\begin{figure}[htb]
 \begin{center}
 \includegraphics[scale=0.8]{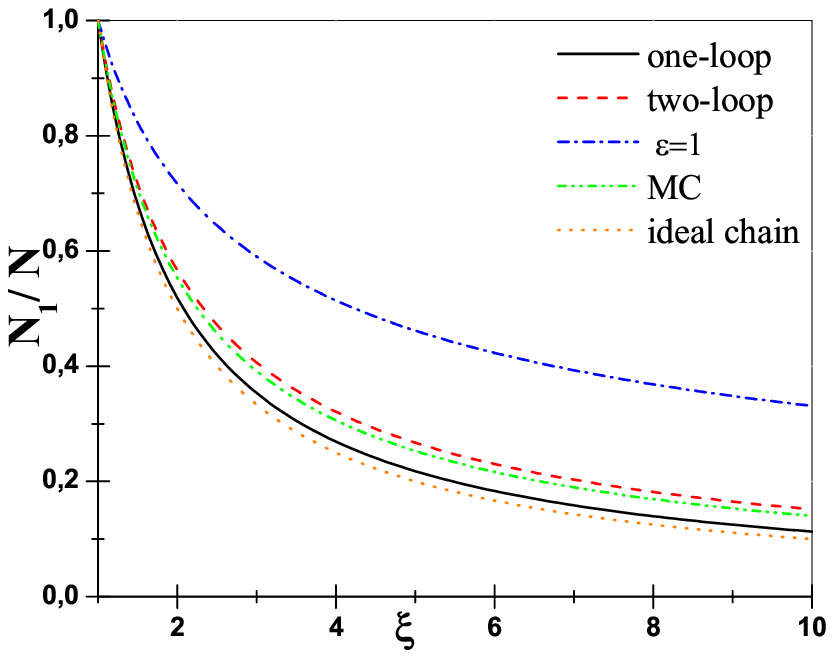}
\end{center}
\caption{Color online. The dependence of the fraction of monomers at
the surface $N_{1}/N$ on the thickness of the adsorbed layer $\xi/l$
for $c<0$ (i.e. below the threshold). Both quantities are
dimensionless.} \label{fig5}
\end{figure}

\begin{figure}[htb]
\begin{center}
\includegraphics[scale=0.9]{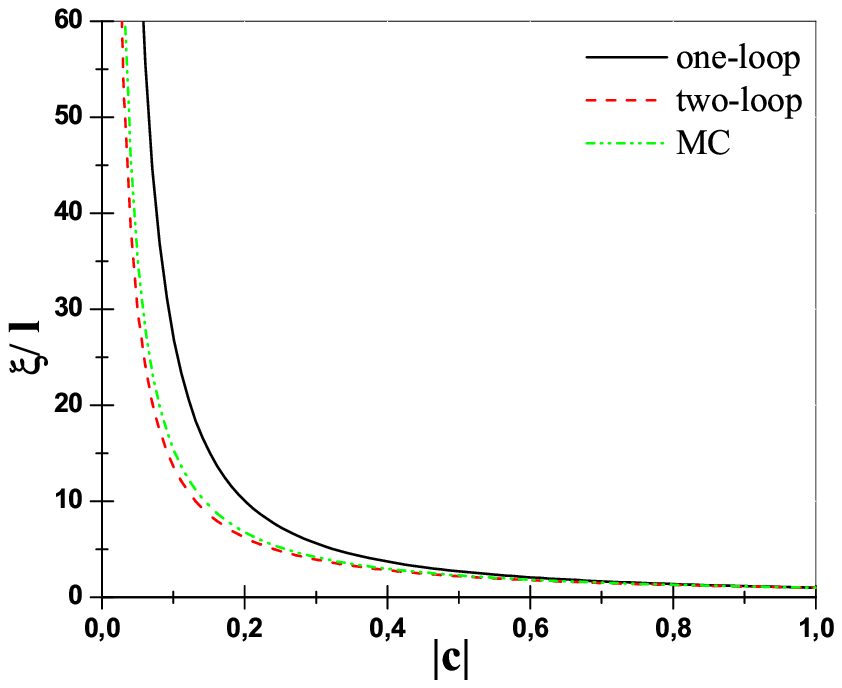}
\end{center}
\caption{Color online. The dependence of the thickness of the
adsorbed layer $\xi/l$ on $|c|$, for $c<0$, where $c\propto
(T-T_a)/T_a$ is the reduced temperature distance from the threshold.
Both quantities are dimensionless. }
 \label{fig6}
\end{figure}
 The scaling behavior is also obeyed by the mean number of the
free ends in the layer between $z$ and $z+dz$, which is proportional
to the partition function of a chain with one end fixed at ${\bf
x}_{A}=({\bf r}_{A},z)$ and the other end free, $Z_{N}(z)$, where

\be Z_{N}(z)= \int_{0}^{\infty}dz^{'}Z^{\lambda}_{N}(z^{'},z).\ee
 The
density of monomers in a layer  at the distance $z$ from the wall to
which one end of the polymer is attached, $M_{N}^{\lambda}(z)$
 scales according to Eq. (\ref{scal}) as well.  For
the above quantities the exponent $a_Y$ in (\ref{scal}) is
$\gamma-1$ and $\gamma_1-\nu$ respectively. Short-distance behavior
($l\ll z\ll \xi_R$) of the two quantities right at the threshold
($c=0$) is \be \label{ZN} Z_{N}(z) \sim z^{a^{'}} N^{b^{'}} \ee
 and
\be \label{MN} M_{N}^{\lambda}(z)\sim z^{-\bar{a}}N^{\bar{b}}, \ee
where $a^{'}=(\gamma-\gamma_{1})/ \nu$, $b^{'}=\gamma_{1}-1$ and
$\bar{a}=1-(1-\Phi)/\nu$, $\bar{b}=-1+\Phi+\gamma_{1}$.

We have obtained corresponding series expansion for the exponent
$\bar{b}$ \be \bar{b}=\frac{1}{2}-\frac{1}{3}( ln
2-\frac{9}{8})\bar{u}+\frac{2}{3} R(0)\bar{u}^{2}, \label{br}\ee
where \be R(0)=-\frac{127}{648}-\frac{3}{2}A+B-\frac{ln 2}{2}
(\frac{3}{2}ln 2-1).\ee We have performed Pade-analysis of the
obtained series. The results are presented in Table 2.

\begin{table}[htb]
\caption{\label{tab:tab2}Pade-analysis of the critical exponents
$(1-\Phi)/\nu$ and $\bar{b}=-1+\Phi+\gamma_{1}$ at the special
transition $c=c_a$ (the case of adsorption at marginal surface) for
$d=3$ up to two-loop order calculated at the pure fixed point
$(u^{*}=1.632$.}
\begin{center}
\begin{tabular}{rrrrrrrr}
\hline $  exp  $~&~$  [0/0] $~&~$  [1/0]  $~&~$  [0/1] $~&~$ [2/0]
$~&~$ [0/2]  $~&~$ [1/1] $~&~$ f_{b} $ \\
 \hline
 $(1-\Phi)/\nu$  &  1.0  &  0.954 &  0.956  & 0.551  &  0.689  & 0.739 & 0.620  \\

 $\bar{b}$ &  1/2  &  0.676  &  0.714  &  0.841 & 0.949 & 0.841 &  0.895  \\

 $\bar{a}$  &  0  &  0.046  &  0.044 & 0.449 & 0.311 & 0.261 &  0.380  \\

\end{tabular}
\end{center}
\end{table}

Figure 7 presents the scaling behavior of the  partition function of
a chain with one end fixed at $({\bf{x}}=({\bf{r}}_{A},z))$  and the
other end free, $Z_{N}$ which is proportional to the mean number of
the free ends in the layer between $z$ and $z+dz$.  Figure 8 shows
the scaling behavior of the density of monomers in a layer at the
distance $z$ from the surface to which one end of the polymer is
attached, $M_{N}^{\lambda}(z)$. To compare the reliability of the
obtained results we showed in Figures 7 and 8 the results obtained
in the frames of $\epsilon$-expansion \cite{EKB2,DD81,DD83} and
Monte Carlo analysis \cite{HG94}.

\begin{figure}[htb]
\begin{center}
\includegraphics[scale=0.9]{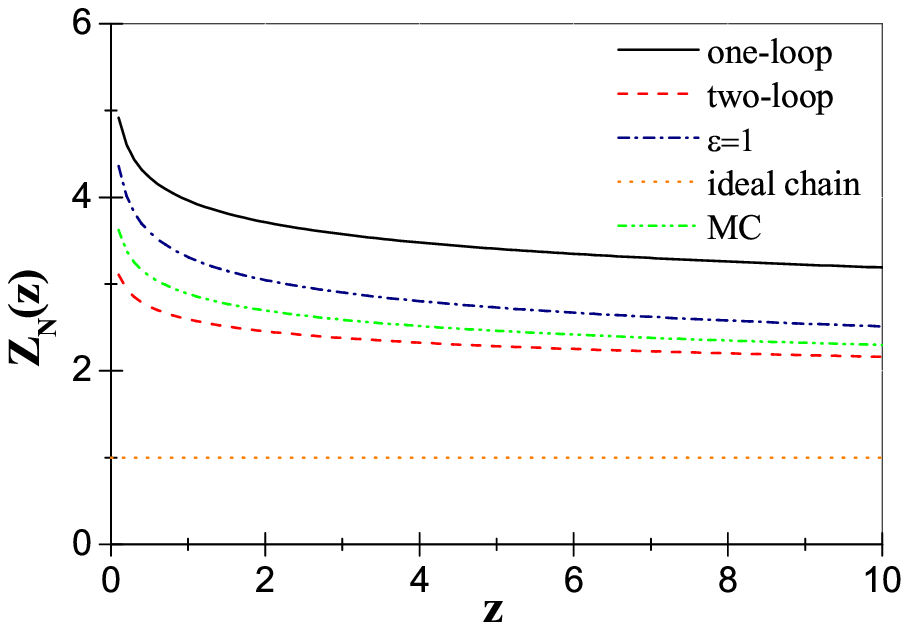}
\end{center}
\caption{Color online. The  partition function $Z_{N}(z)$  just at
the adsorption
 thereshold $c=0$ and for $N=100$, as a function of $z/l$ for $1\ll
 z/l\ll N^{\nu}$.  $Z_{N}(z)$ is
 dimensionless and $l$ is the microscopic length scale. }
 \label{fig7}
\end{figure}

\begin{figure}[htb]
\begin{center}
\includegraphics[scale=0.8]{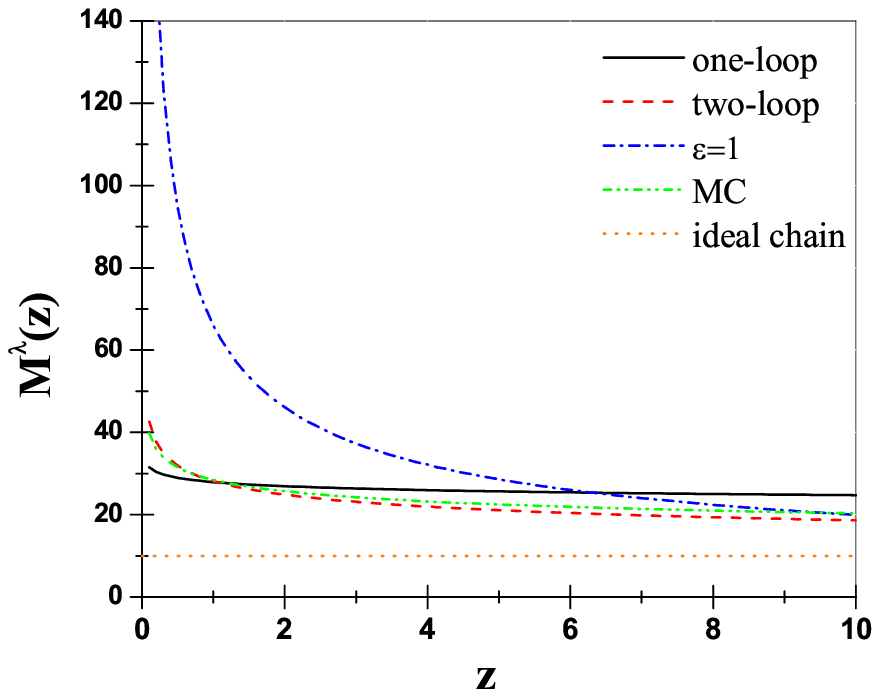}
\end{center}
\caption{Color online. The  density of monomers $M^{\lambda}(z)$ in
the layer at the
 distance $z$ from the surface to which one end of the chain is
 attached for $1\ll z/l \ll N^{\nu}$ just at the threshold $c=0$ and for
 $N=100$. $M^{\lambda}(z)$ is in arbitrary units and $z/l$ is
 dimensionless.}
\label{fig8}
\end{figure}

\section{Critical behavior of long-flexible polymer chains near the repulsive surface
for $c>c_{0}^{sp}$ (ordinary transition) }

In the case of repulsive surface the spin correlation functions with
one and two points at the surface are simply

\be G^{\lambda}(z,c_{0})\sim \frac{1}{c}
z^{1-\eta^{ord}_{\perp}}G_{\lambda}(\tau z^{1/\nu})\ee
\begin{equation}
G_{\parallel,\perp}(r;\mu_{0},u_{0},c_{0})\sim \frac{1}{c}
r^{-(d-2+\eta_{\parallel,\perp}^{ord})}G_{\parallel,\perp}(\tau
r^{1/\nu}). \label{29c}
\end{equation}
The corresponding short-distance behavior $1\ll z/l,r/l\ll N^{\nu}$
 of the partition functions
$Z^{\lambda}(0,z)\sim z^{a_{\lambda}} N^{b_{\lambda}}$ and
$Z^{\parallel,\perp}(r)\sim r^{a_{\parallel,\perp}}
N^{b_{\parallel,\perp}}$ have the following critical exponents:
$a_{\lambda}= \eta_{\parallel}^{ord}-\eta_{\perp}^{ord}$,
$b_{\lambda}=-1+\gamma_{\parallel}^{ord}$ and
$a_{\parallel,\perp}=2-\eta_{\parallel,\perp}^{ord}$,
$b_{\parallel,\perp}=-1-\nu d$.

The series expansions for the critical exponent
$\eta_{\parallel}^{ord}$  up to two-loop order was obtained
previously for the pure case in \cite{DSh98} and for the case with
some amount of disorder in \cite{UShCh01}(the pure case can be
obtained from the formulas (4.13) in \cite{UShCh01} in the limit
$n\to 0$ and at $v=0$). Thus, we have \be
\eta_{\parallel}^{ord}=2-\frac{u}{8}-\frac{3}{4}C(0)u^{2},\label{etaord}
\ee where $C(0)=C+\frac{7}{48}$ and the constant $ C\simeq{107\over
162}-  {7\over 3} \ln{4\over 3}-0.094299\simeq -0.105063$ stems from
two-loop contribution. Table 3 represents the obtained one-loop and
two-loop calculations for the corresponding critical exponents
characterizing the behavior of long-flexible polymer chains near the
repulsive surface.

\begin{table}[htb]
\caption{\label{tab:tab3}Critical exponents characterizing scaling
critical behavior of the long-flexible polymer chains near the
repulsive surface $c>c_{0}^{a}$ for $d=3$ up to one-loop and
two-loop order calculated at the pure fixed point $u^{*}=1.632$.}
\begin{center}
\begin{tabular}{rrrrrrrrrrr}
\hline $    $~&~$ a_{\lambda} $~&~$ b_{\lambda} $~&~$ a_{\parallel}
$~&~$ b_{\parallel} $~&~$ a_{\perp} $~&~$ b_{\perp} $~&~$ a^{'}
$~&~$ b^{'}$~&~$ \bar{a} $~&~$ \bar{b} $ \\
\hline one-loop & 0.903 & -1.500  & 0.195 & -2.824 &
1.098 & -2.824 & -0.926 & -0.333 & -0.645 & -0.333 \\
\hline
 two-loop &  0.817  &  -1.388  &  0.340  & -2.764  &  1.157  & -2.764 &
 -0.819
 & -0.320 &  -0.701  &  -0.320   \\

\end{tabular}
\end{center}
\end{table}

Short-distance behavior ($l\ll z\ll \xi_R$) of the quantities
$Z_{N}(z)$ and $M_{N}^{\lambda}(z)$ near the repulsive surface
($c>0$) is characterized by the new exponents
$a^{'}=(\gamma-\gamma_{1}^{ord})/ \nu$, $b^{'}=\gamma_{1}^{ord}-1$
and $\bar{a}=1-1/\nu$, $\bar{b}=-1+\gamma_{1}^{ord}$. The results of
calculations for one-loop and two-loop order for these exponents are
presented in Table 3. Figures 9-11 present the scaling behavior of
the corresponding partition functions, $Z^{\perp}(z),
Z^{\parallel}(r), Z_{N}(z)$, and Figure 12 shows the scaling
behavior of the density of monomers in a layer at the distance $z$
from the repulsive surface (for $c>c_{0}^{sp}$) to which one end of
the polymer is attached, $M_{N}^{\lambda}(z)$. For comparison of the
reliability of the obtained in the frames of RG field theory results
at fixed dimension $d=3$ we also show in Figures 8-12 results
obtained from $\epsilon$- expansion \cite{EKB82}. Besides, Figure 12
contains results obtained by recent Monte Carlo calculations
\cite{HG94}. We see, that the two-loop order result of RG field
theory at fixed dimensions $d=3$ gives more reliable estimates of
the surface critical exponents and scaling critical behavior than
$\epsilon$-expansion series.

\begin{figure}[htb]
 \begin{center}
 \includegraphics[scale=0.8]{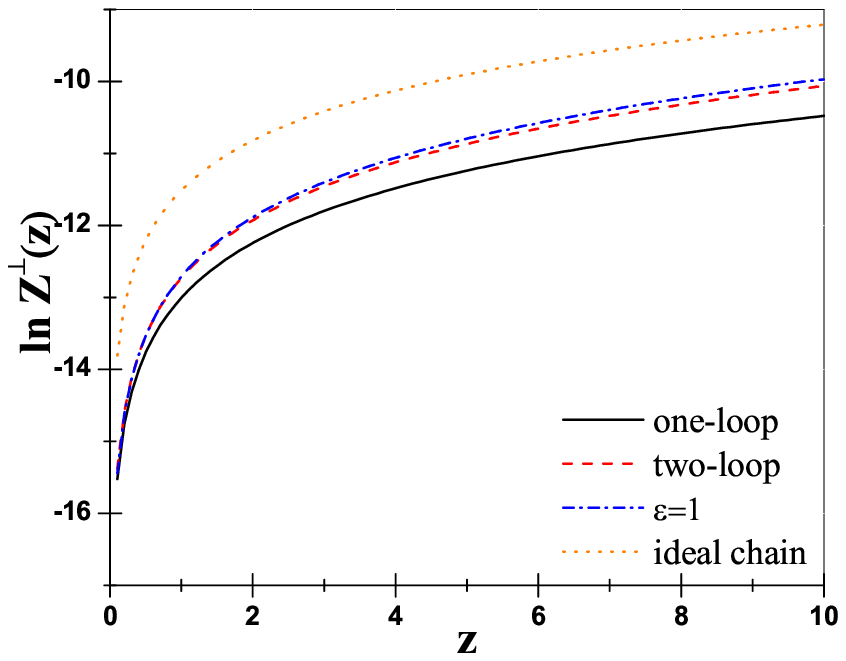}
\end{center}
\caption{Color online. The  partition function $ln Z^{\perp}(z)$ at
the ordinary transition $c>0$ (repulsive surface) and for $N=100$,
as a function of $z/l$ for $1\ll z/l\ll N^{\nu}$. Both values
dimensionless. }\label{fig9}
\end{figure}
\begin{figure}[htb]
 \begin{center}
 \includegraphics[scale=0.8]{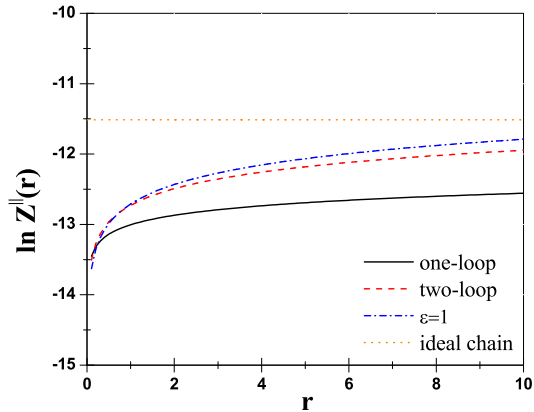}
\end{center}
\caption{Color online. The  partition function $ln Z^{\parallel}(r)$
at the ordinary transition $c>0$ (repulsive surface) for $N=100$, as
a function of $r/l$ for $1\ll r/l\ll N^{\nu}$. Both values
dimensionless.
 }
 \label{fig10}
\end{figure}

\begin{figure}[htb]
\begin{center}
\includegraphics[scale=0.9]{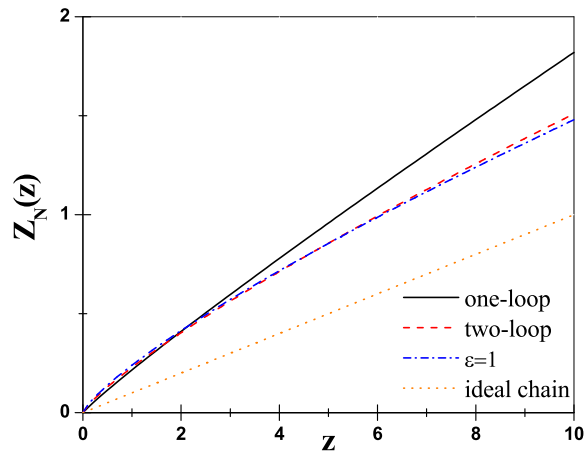}
\end{center}
\caption{Color online. The  partition function $Z_{N}(z)$ at the
ordinary transition $c>0$ (repulsive surface)
 and for $N=100$, as a function of $z/l$ for $1\ll
 z/l\ll N^{\nu}$.  $Z_{N}(z)$ is
 dimensionless and $l$ is the microscopic length scale. }
 \label{fig7}
\end{figure}

\begin{figure}[htb]
\begin{center}
\includegraphics[scale=0.8]{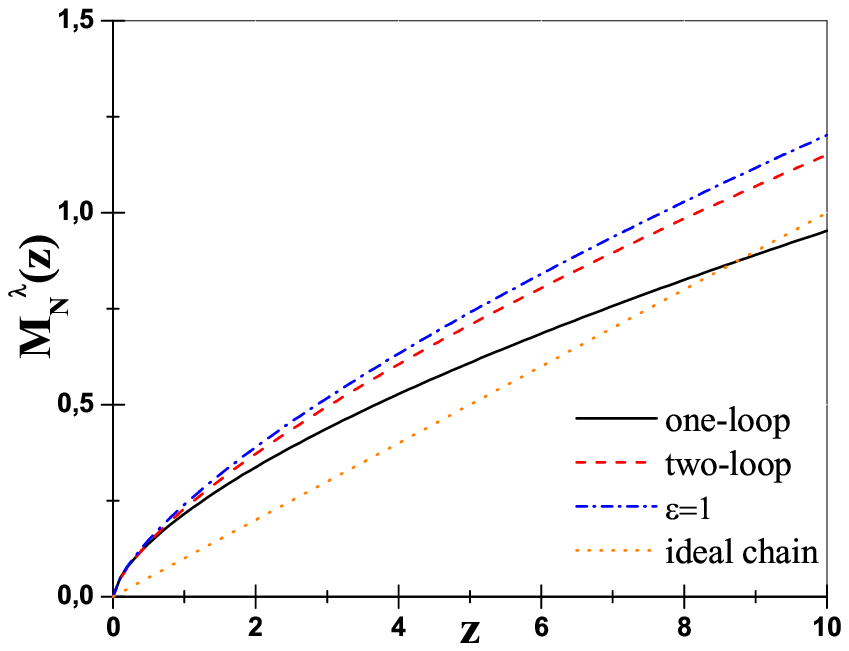}
\end{center}
\caption{Color online. The  density of monomers $M^{\lambda}(z)$ in
the layer at the
 distance $z$ from the surface to which one end of the chain is
 attached for $1\ll
 z/l\ll N^{\nu}$ at the ordinary transition $c>0$ (repulsive surface) and for
 $N=100$. $M^{\lambda}(z)$ is in arbitrary units and $z/l$ is
 dimensionless.}
\label{fig12}
\end{figure}

\section{Center-adsorbed star polymers}

The above mentioned scheme is possible to apply to the scaling
analysis of center-adsorbed star polymers with $f$ arms. We consider
a network containing  $f$  long-flexible linear polymer chains which
have the same length. One of the most important characteristics for
the center-adsorbed star polymers is the center-end distribution
function, $g(r^{CE}_{\parallel},z^{E})$, ( where
$r^{CE}_{\parallel}=|r^{E}_{\parallel}-r^{C}_{\parallel}|$ is the
center-end distance parallel to the surface and $z^{E}$ denotes the
distance of the end point from the surface) which at short distances
is described by nontrivial exponents \cite{Ochno1}. As was indicated
in \cite{Ochno1}, for center-adsorbed stars at repulsive or marginal
surface the center-end distribution function $g(r^{CE},z)$ behaves
as \be g({\bf{r}}_{\parallel}^{CE},0)\sim
(r^{CE}_{\parallel})^{\theta_{\parallel}(f)} \ee and \be
g(0,z^{E})\sim (z^{E})^{\theta_{\perp}(f)}, \ee with exponents \be
\theta_{\parallel}(f)=(\gamma_{1}-\gamma_{s}(f+1)+\gamma_{s}(f)-1)/\nu,
\ee \be
\theta_{\perp}(f)=(\gamma-\gamma_{s}(f+1)+\gamma_{s}(f)-1)/\nu, \ee
where $\gamma_{1}$ is the surface critical exponent of a linear
polymer chain with one end at the surface. The Pade-Borel analysis
of the corresponding series for the critical exponent $\gamma_{1}$
up two-loop order at fixed dimensions $d=3$ was obtained by
\cite{DSh98}. The critical exponents of star polymers$\gamma (f)$
and $\gamma_{s}(f)$ are \be
\gamma(f)=\alpha-1+(\gamma-\alpha)f/2+\Delta_{f}, \ee and \be
\gamma_{s}^{sp}(f)=\alpha-1+\nu+(\gamma-\alpha)f/2+
\Delta_{f}^{sp},\ee where exponent $\alpha$ can be obtained via
hyperscaling relation $\alpha=2-\nu d$. The first several exponents
$\Delta$'s can be expressed  via usual critical exponents of linear
polymer chain. As was noted in Ref.\cite{Ochno1},
$$ \Delta_{1}=1+(\gamma-\alpha)/2, \quad\quad \Delta_{2}=1$$
$$ \Delta_{1}^{'}=(d-1)\nu/2+\gamma_{\parallel}/2, \quad\quad
\Delta_{2}^{'}=-\nu$$,
$$\Delta_{1}^{sp}=(d-1)\nu/2+\gamma_{11}^{sp}/2.\quad\quad
\Delta_{2}^{sp}=\Phi.$$

We have performed calculation of the critical exponents of
center-adsorbed star polymers up to one-loop and two-loop order on
the basis of the best estimates for surface critical exponents
obtained in the frames of RG field theory at fixed dimensions
\cite{DSh98}. The results of our calculations are presented in Table
4.
\begin{table}[htb]
\caption{\label{tab:tab4}Critical exponents characterizing scaling
critical behavior of the center-adsorbed  star polymer chains for
$d=3$ up to one-loop and two-loop order.}
\begin{center}
\begin{tabular}{rrrrrrrr}
\hline $    $~&~$ \Delta_{1} $~&~$ \Delta_{1}^{'} $~&~$
\Delta_{1}^{sp} $~&~$ \gamma(1) $~&~$ \gamma(2) $~&~$
\gamma_{s}^{sp}(1) $~&~$ \gamma_{s}^{sp}(2) $\\
\hline one-loop & 1.527 & 0.358  & 0.973 & 1.230 &
1.230 & 1.284 & 1.262 \\
\hline
 two-loop &  1.463  &  0.394  &  0.921  & 1.162  &  1.162  & 1.260 &
 1.268   \\

\end{tabular}
\end{center}
\end{table}
The knowledge of the critical exponents $\gamma(f)$ and
$\gamma_{s}^{sp}(f)$ gives access to the critical exponents
$\theta_{\parallel}(f)$ and $\theta_{\perp}(f)$ and allows to obtain
the configuration-number exponent $\gamma_{\it{G}}$ for a polymer
network $\it{G}$ with $n_{j}$ $j$-functional units in the bulk and
$n_{j}^{'}$ $j$-functional units at the surface and with total
number of linear polymer chains $f$ with the same length.  According
to the \cite{Ochno1}, for $\gamma_{\it{G}}$ the following scaling
relation are valid \be
\gamma_{\it{G}}=\alpha-1-f+\nu+\sum_{j=1}^{\infty}
(n_{j}\Delta_{j}+n_{j}^{'}\Delta_{j}^{sp}), \ee

where $\Delta_{j}$ and $\Delta_{j}^{sp}$  $(j=1,2)$ was described
early for the case of center-adsorbed star polymer chains at the
marginal surface (special transition). Exactly the same scaling
relation takes place for the case of star polymer chains near the
repulsive surface, but in this case $\Delta_{j}^{sp}$ must be
replaced by $\Delta_{j}^{'}$.

\section{Conclusions}

In the present paper, the RG field-theoretical analysis at the fixed
dimensions $d=3$ of adsorption of long-flexible polymer chains with
excluded volume interactions at repulsive, marginal and adsorbing
surface was performed. The scaling functions do not depend on
short-range correlations among the monomers along the polymer chain
and precise form of the monomer-surface interaction and the value
$T_{a}$. The particular attention was payed to the region around
adsorption threshold $c\propto(T-T_a)/T_a)$, where polymer chains
conformation changes from nonadsorbed three-dimensional state
(ordinary transition) to two-dimensional one (extraordinary
transition). We have performed scaling analysis of the surface
correlation functions and the corresponding partition function of a
chain with one end at the surface and another one free and partition
function with two ends at the surface. We have analyzed the parallel
and perpendicular parts of the end-to-end distance
$<R_{\parallel}^{2}>^{1/2}$ and $<R_{\perp}^{2}>^{1/2}$.  The
behavior of the average end-to-end distance at and above the
threshold is independent of the surface critical exponents. We also
have performed analysis of the fraction of monomers of the polymer
chains attached to the repulsive, marginal and attractive surfaces.
Right at and above threshold the average number of polymer free ends
and the average number of monomers depend on the distance from the
 surface. The distribution of monomers at different distances from the
 surface at and above the adsorption threshold, as well as the crossover
 behavior to the adsorbed state, can only be determined with the help
 of the surface critical exponents.
We have performed our scaling analysis on the basis of RG field
theory approach at fixed dimensions $d=3$. We have found, that the
two-loop order results of RG field theory at fixed dimensions $d=3$
give more reliable estimates of the surface critical exponents and
scaling critical behavior than previous theoretical results obtained
on the basis of $\epsilon$-expansion. The obtained two-loop order
results more better consist with recent Monte Carlo calculations
\cite{ML88,HG94}.

Besides, we have performed scaling analysis of the center-adsorbed
star polymers with $f$ arms of the same length. We have obtained
correspondent critical exponents $\gamma(f)$ and
$\gamma_{s}^{sp}(f)$ up to one-loop and two-loop order at fixed
dimensions $d=3$, which allow to describe the center-end
distribution function,$g(r^{CE},z^{E})$, and the
configuration-number exponent $\gamma_{G}$ for a polymer network
$\it{G}$ with $n_{j}$ $j$-functional units in the bulk and
$n_{j}^{'}$ $j$-functional units at the surface and with total
number of linear polymer chains $f$ with the same length. More
detailed discussion of the behavior of center-adsorbed star polymer
chains will be presented in forthcoming paper.

\section*{Acknowledgments}
 I should like to thank Dr. M.Shpot and Prof. A.Ciach for a useful discussion.

\end{document}